# The Solving of the Problems with Random Division of an Interval with Use of Computer Analytic Programs

Aleksandr Reznik, Vitaly Efimov, Aleksandr Soloviev, and Andrey Torgov

*Institute of Automation and Electrometry, Siberian Branch, Russian Academy of Sciences, pr. Akademika Koptyuga 1, Novosibirsk, 630090 Russia*

Corresponding author : A.Reznik

E-mail: reznik@iae.nsk.su

**Abstract**: An original approach to solving rather difficult probabilistic problems arising in studying the readout of random discrete fields and having no exact analytical solutions at the moment is proposed. Several algorithms for direct, iterative, and combinatorial-recursive calculations of multidimensional integral expressions, which can describe partial solutions of these problems, are presented (these solutions are further used to search for the common closed analytical regularities). The huge volume of necessary calculations forced us to formalize completely the algorithms and to transfer all the burden of routine analytical transforms to a computer. The calculations performed helped us to establish (and to prove later) a number of new earlier unknown probabilistic formulas responsible for random division of an interval. Thus, we managed to realize (which is extremely rare in scientific practice) the idea proposed by John von Neumann: if researchers face difficult scientific problem and don't know how to find its exact analytical solution, they use computer calculations to find the widest range of partial solutions directing to correct "answer", and then prove its validity. One more important feature of this study is the fact that we introduced a new concept of "three-dimensional generalized Catalan numbers" and found their explicit form in solving problems associated with random division of an interval.

## 1. Introduction

Our research on the reliability of readout of random discrete-point fields led us to the following very simple (in definition) probabilistic problem:

«Let $n$ points $x_1, x_2, \ldots, x_n$ be randomly dropped on an interval $(0,1)$, i.e., there are $n$ independent tests of a random variable uniformly distributed in the interval $(0,1)$. It is required to determine the probability $P_{n,k}(\varepsilon)$ of an event that there is no subinterval $\Omega_\varepsilon \subset (0,1)$ of length $\varepsilon$ containing more than $k$ points».

The apparent simplicity of this problem is illusory, and its analytic solution is known [12-13] only for $k=1$:

$$P_{n,1}(\varepsilon) = (1-(n-1)\varepsilon)^n, \quad (0 \leq \varepsilon \leq 1/(n-1)). \qquad (1)$$

It should be noted that many problems that involve random division of an interval [10] are simple in statement, but have no exact analytical solution. It is an example we have here.

One way to find solution (1) is to represent the probability $P_{n,1}(\varepsilon)$ as a repeated integral

$$P_{n,1}(\varepsilon) = n! \int_{(n-1)\varepsilon}^{1} dx_n \left\{ \int_{(n-2)\varepsilon}^{x_n-\varepsilon} dx_{n-1} \cdots \left\{ \int_{2\varepsilon}^{x_4-\varepsilon} dx_3 \left\{ \int_{\varepsilon}^{x_3-\varepsilon} dx_2 \left\{ \int_{0}^{x_2-\varepsilon} dx_1 \right\} \right\} \right\} \right\}. \qquad (2)$$

Successive integration of Eq. (2) with respect to the variables $x_1, x_2, ..., x_n$ yields relation (1). Unfortunately, for $k>1$, the probability $P_{n,k}(\varepsilon)$ cannot be reduced to a single repeated integral; therefore, the algorithmic difficulties that have to be overcome to find the probability $P_{n,k}(\varepsilon)$ grow to such an extent that today there is no exact analytic solution of this problem even for $k=2$. The present paper describes possible approaches to solving the above-discussed problem.

## 2. Computer-aided analytical calculation of probabilistic formulas

In general, the probability $P_{n,k}(\varepsilon)$ can be represented in the form

$$P_{n,k}(\varepsilon) = n! \int \cdots \int_{D_{n,k}(\varepsilon)} dx_1 \ldots dx_n, \qquad (3)$$

where the domain of integration $D_{n,k}(\varepsilon) \subset R^n$ is described by a system of linear inequalities

$$\begin{cases} 0 < x_1 < x_2 < \ldots < x_{n-1} < x_n < 1, \\ x_{k+1} - x_1 > \varepsilon, \\ x_{k+2} - x_2 > \varepsilon, \\ \vdots \\ x_n - x_{n-k} > \varepsilon. \end{cases} \qquad (4)$$

Further, integral (3) can be written in equivalent form:

$$P_{n,k}(\varepsilon) = n! \int_{-\infty}^{+\infty} \cdots \int_{-\infty}^{+\infty} 1[x_1] 1[x_2] \ldots 1[x_n - x_{n-1}] 1[1 - x_n] 1[x_{k+1} - x_1 - \varepsilon] \times$$

$$\times 1[x_{k+2} - x_2 - \varepsilon] \ldots 1[x_n - x_{n-k} - \varepsilon] dx_1 \ldots dx_n, \qquad (5)$$

where $1[z] = \begin{cases} 0, & z \leq 0, \\ 1, & z > 0. \end{cases}$ Then, the *n*-dimensional integral (5) is converted to a set of repeated integrals with already set limits of integration by means of successive application of the relation

$$\left(\prod_{j=1}^{l} 1[x_r - \alpha_j]\right)\left(\prod_{i=1}^{m} 1[\beta_i - x_r]\right) = $$
$$= \sum_{j=1}^{l}\sum_{i=1}^{m} 1[x_r - \alpha_j]1[\beta_i - x_r]1[\beta_i - \alpha_j]\left(\prod_{\substack{q=1 \\ q \neq j}}^{l} 1[\alpha_j - \alpha_q]\right)\left(\prod_{\substack{s=1 \\ s \neq i}}^{m} 1[\beta_s - \beta_i]\right). \quad (6)$$

(The form of Eq. (6) means that the expressions $\alpha$ and $\beta$ do not contain the variable $x_r$). The algorithm (3)-(6) allows a constructive calculation of the formulas for $P_{n,k}(\varepsilon)$ for fixed values of *n* and *k*. The main difficulty of application of this procedure lies in the fact that for it is practically impossible to perform manually all calculations necessary for settling the limits of integration, verifying the correctness of all intermediate systems of inequalities, and direct calculating of the repeated integrals for $n>4$. For these reasons, we created a program package based on algorithm (3)-(6) and performing all analytical calculations [6].

As one more approach to finding particular solutions of the initial problem, we proposed the following procedure. By analogy with the known solution (1) valid for *k*=1, we tried to find a general solution $P_{n,2}(\varepsilon)$ for *k*=2. Contrary to the above-described algorithm, we use an absolutely different mathematical technique. With the aid of purely combinatorial tools, we compose a recursive algorithm where the formulas for $P_{n,2}(\varepsilon)$ as functions of the continuous argument $\varepsilon$ are obtained from a discrete-combinatorial scheme by passing to the limit. We used the following discrete-combinatorial model. The interval (0,1) is considered to be divided into *r* equal quantums. Random throwing of *n* points on the interval (0,1) is interpreted as random throwing of *n* indistinguishable balls into *r* boxes. A set of *l* adjacent quantums serves as an analog of a subinterval of length $\varepsilon$. The outcome of throwing when none of such *l*-subintervals inside the initial *r*-interval (0,1) has more than two points is considered as "successful," and the ratio of the total number of "successful throws" $Q(r,n,l)$ to the overall number of possible outcomes of throwing $Q(r,n)$ is taken as a discrete-combinatorial analog for the probability $P_{n,2}(\varepsilon)$. If it were possible to find a closed analytical expression for $Q(r,n,l)$, it would be equivalent to a complete solution of the problem for the case with *k*=2 because

$$P_{n,2}(\varepsilon) = \lim_{\substack{r \to \infty \\ (l/r) \to \varepsilon}} \frac{Q(r,n,l)}{Q(r,n)},$$

and the expression for $Q(r,n)$ is known [11]:

$$Q(r, n) = C_{n+r-1}^{r-1} = \frac{(n+r-1)!}{n!(r-1)!}$$

It does not seem possible to obtain exact and analytically closed (relative to the parameters *r*, *n,* and *l*) formulas for $Q(r,n,l)$, but we managed to construct recursive relations for their consecutive calculation. It is important to note that these recursive relations are very cumbersome, and that is why we do not present them in this work. In view of the large volume of routine analytic calculations necessary to perform such recursion, the entire algorithm of combinatorial computations was again realized as a computer program.

Our third software package is based on multiple differentiation of the initial integral (3) with respect to the parameter $\varepsilon$ and further reconstruction of the formulas for $P_{n,k}(\varepsilon)$ with the values of all derivatives $\frac{d^{(j)}P(\varepsilon)}{d\varepsilon^{(j)}}$, ( $j = 0,1,..., n$ ) at zero (i.e., at $\varepsilon=0$). The main advantage of this algorithm is the fact that its implementation allows us to replace labor-consuming procedures of determining the integration limits and subsequent multidimensional integration by elementary operations of substitution and replacement of variables. In this case, it becomes possible owing to application of the obvious equalities

$$\frac{d}{dz}1[z] = \delta(z); \quad \int_{-\infty}^{+\infty} \delta(z)F(z)dz = F(0),$$

which were used to calculate integral (5) (here δ (z) is the Dirac delta function). Application of this algorithm, however, is limited because it allows calculating the formulas for $P_{n,k}(\varepsilon)$ only in a single range of variation of the parameter $\varepsilon$ adjacent to zero.

Using three above-mentioned program systems, we evaluated the formulas for $P_{n,k}(\varepsilon)$ for particular values of *n* and *k* (*k*<*n*) up to *n*=14 and for all ranges of variation of the parameter $\varepsilon$. Further these computer analytical calculations helped us to establish at first, and then to prove strictly the new earlier unknown probabilistic regularities relating to a random division of an interval.

## 3. Proof of "computer" formulas with the use of the Catalan numbers

The analysis of the formulas for $P_{n,k}(\varepsilon)$ calculated on a computer allows us to establish a number of new previously unknown analytical dependences. In particular, for even values of $n=2m$ and $k=2$, we established the formula

$$P_{2m,2}(\varepsilon) = \frac{1}{m} C_{2m}^{m-1} (1-(m-1)\varepsilon)^{2m}, \qquad (7)$$

which is valid if $1/m < \varepsilon < 1/(m-1)$. The coefficients $(1/m) C_{2m}^{m-1}$ in Eq. (7) are the classical Catalan numbers known from Leonard Euler's works, but they are still of interest [1,4,9] because they form the basis of enumerative combinatorics [14]. It is curious to note that relation (7) was "prompted" by a computer and was published as a scientific hypothesis more than thirty years ago [5], and a strict mathematical proof of this formula was obtained rather recently [7]. Thereby, we realized in practice the classical advice of John von Neumann: if you cannot find a straightforward solution of a difficult scientific problem, try to perform laborious auxiliary calculations on a computer. If you are lucky, these auxiliary computer calculations can "prompt" you the right answer, which you will prove further.

Recently we managed to prove that the probability $P_{n,k}(\varepsilon)$ for $k=2$ and odd values of $n=2m+1$ is represented as

$$\begin{aligned} P_{2m+1,2}(\varepsilon) = & \; C_{2m+1}^{m+1}(1-m\varepsilon)^{m+1}(1-(m-1)\varepsilon)^{m} - \\ & - 2 C_{2m+1}^{m+2}(1-m\varepsilon)^{m+2}(1-(m-1)\varepsilon)^{m-1} + \\ & + C_{2m+1}^{m+3}(1-m\varepsilon)^{m+3}(1-(m-1)\varepsilon)^{m-2}, \end{aligned} \qquad (8)$$

if $1/(m+1) < \varepsilon < 1/m$.

It turned out that it is much more difficult to find and mathematically substantiate relation (8) than to prove formula (7). Thus, one stage of this proof forced us at first to introduce a new concept of "three-dimensional generalized Catalan numbers" and next to calculate their explicit form. In our investigations, these numbers appeared when we had to find the total number of specific permutations of elements of three subsets, with each of these subsets being represented as a ranked sequence of uniformly distributed random variables.

If we put aside the researches dealing with random division of an interval, then the problem leading to three-dimensional generalized Catalan numbers can be formulated in the following most transparent form: "It is necessary to find $Q_{l,m,n}$, which is the exact number of different words of length $(l+m+n)$ that can be formed from $l$ letters "a", $m$ letters "b", and $n$ letters "c" with two conditions being satisfied simultaneously: 1) the number of the letters "b" never exceeds the number of the letters "a" if the word is viewed from left to right; 2) the number of the letters "c" never exceeds the number of the letters "a" if the word is viewed from right to left".

Having reduced this problem with three-letter words to a geometrical problem of searching for paths on a three-dimensional discrete lattice, we managed to show that

$$Q_{l,m,n} = \frac{(l+m+n)!}{l!\,m!\,n!} \times \frac{(l+1)(l+2) - (m+n)(l+2) + mn}{(l+1)(l+2)}, \qquad (9)$$

if $\quad m+n < l+2.$ \hfill (9a)

A detailed proof of this equality can be found in [8]. Here we only briefly describe the essence of the algorithm. We consider (see Fig.) various paths on a three-dimensional discrete lattice in a coordinate system $(X, Y, Z)$, which lead from the point $(0,0,0)$ to the point $(l, m, n)$. Each word is put into correspondence to one of these paths. The letter "$a$" means the motion from the current point $(i, j, k)$ to the neighboring point $(i+1, j, k)$; the letter "$b$" indicates the motion to the point $(i, j+1, k)$, and the letter "$c$" means the motion to the point $(i, j, k+1)$. Our task is to find the number of such paths from the point $(0,0,0)$ to the point $(l, m, n)$ that intersect neither the plane $P_1$ defined by the equation $X-Y=0$ (i.e., all the paths considered here lie in the half-space $X \geq Y$) nor the plane $P_2$ defined by the equation $X-Z+n-l=0$. Thus, each of the paths satisfying both conditions lies not only in the half-space $X \geq Y$, but also in the half-space $l-X \geq n-Z$.

The thus-reformulated problem is solved as follows. From the total number of paths $S = (l+m+n)!/(l!\,m!\,n!)$ leading from the point $(0,0,0)$ to the point $(l, m, n)$, we subtract the number of paths $Q$ that intersect at least one of the planes $P_1$ or $P_2$. In turn, to calculate $Q$, we have to sum up the number $Q_1$ of paths intersecting the plane $P_1$ and the number $Q_2$ of paths intersecting the plane $P_2$, and then subtract from the resultant sum the number $Q_{12}$ of paths intersecting both the plane $P_1$ and the plane $P_2$ (as they are taken into account two times in this sum).

We showed [8] that

$$Q_1 = \frac{(l+m+n)!}{(l+1)!(m-1)!\,n!}; \qquad Q_2 = \frac{(l+m+n)!}{(l+1)!\,m!(n-1)!};$$

$$Q_{12} = \frac{(l+m+n)!}{(l+2)!(m-1)!(n-1)!}.$$

Therefore, we have

$$Q_{l,m,n} = S - Q_1 - Q_2 + Q_{12} = \frac{(l+m+n)!}{l!\,m!\,n!}\left[1 - \frac{m+n}{l+1} + \frac{mn}{(l+1)(l+2)}\right]. \qquad (10)$$

Thus, the proof of equality (9) is finished. If the condition (9a) isn't satisfied (but, inequalities of $m,n \leq l$ are natural, right), the ratio (10) becomes a little more complicated:

$$Q_{l,m,n} = \frac{(l+m+n)!}{l!\,m!\,n!}\left[1 - \frac{m+n}{l+1} + \frac{mn}{(l+1)(l+2)}\right] + \\ + \frac{(l+m+n)!}{(m+n-l-2)!(n+1)!(n+1)!} - \frac{(l+m+n)!}{(m+n-l-2)!\,n!(n+2)!} \quad . \quad (11)$$

$$\text{if} \quad m,n \leq l. \quad (11a)$$

We named the numbers $Q_{l,m,n}$ as "three-dimensional generalized Catalan numbers" bearing in mind that these numbers extend the traditional Catalan sequence known from many applications (see e.g. [2-3]) and obtained from Eq. (9) at $n=0$ and $l=m$. Equation (10) and more general equation (11) are not only useful in solving applied probabilistic and statistic problems, but have also independent theoretical interest.

## 4. Conclusion

To find exact analytical solutions of the challenging probabilistic problems arising in studying reliability of the readout of random discrete-point structures, some program systems for performing labor-consuming analytical calculations are proposed and realized. By means of the developed software packages, it became possible to calculate a wide set of particular solutions, and the further analysis of the obtained "computer" formulas allowed us to establish (and later on to prove strictly) a number of new earlier unknown analytical equalities. These formulas are necessary to solve many problems dealing with random division of an interval.

One more distinctive feature of the presented work is introduction of a new concept of "three-dimensional generalized Catalan numbers" for the investigations to be successful. Having allocated this issue as a separate subtask, we managed to find a simple and transparent interpretation of this generalization of the classical Catalan sequence (as a solution of a problem with special three-letter words). We found the explicit form of the three-dimensional generalized Catalan numbers by means of reducing this new "linguistic" subtask to a geometrical problem of searching for special paths on a discrete three-dimensional lattice under certain constraints. The subtask with the three-dimensional generalized Catalan numbers has also independent theoretical significance.

# Acknowledgements

This work was supported by the Russian Foundation for Basic Research (Project No. 13-01-00361), by the Presidium of the Russian Academy of Sciences (Project No. 11/2012), and by the Siberian Branch of the Russian Academy of Sciences (Integration Project SB RAS and NASB No. 16/2012).

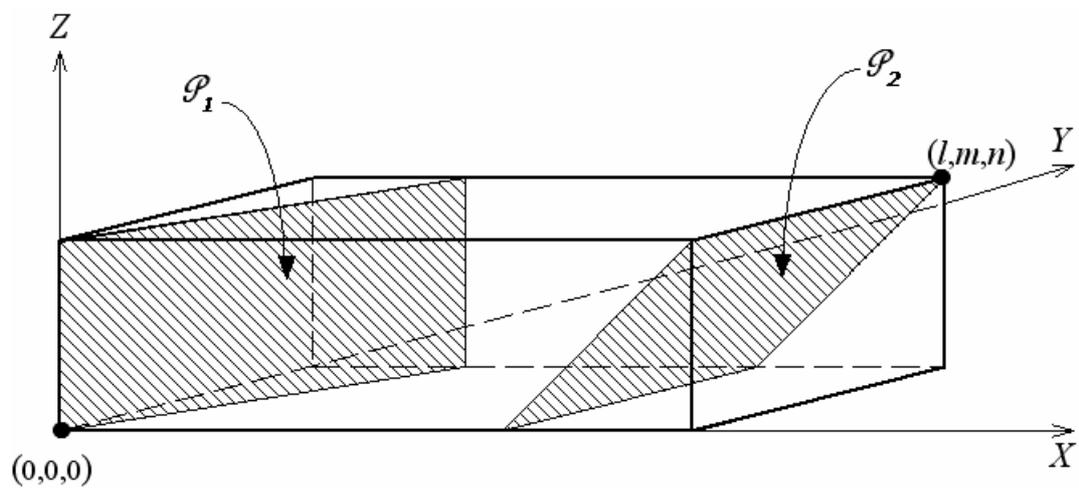

**Fig.** Parallelepiped with bounding planes **P**$_1$ and **P**$_2$.